\documentclass[12pt]{iopart}
\usepackage{graphicx}
 \newcommand {\be} {\begin{equation}}
\newcommand {\bea} {\begin{eqnarray} \nonumber }
\newcommand {\ee} {\end{equation}}
\newcommand {\eea} {\end{eqnarray}}

\newcommand {\bc} {\begin{center}}
\newcommand {\ec} {\end{center}}
\newcommand {\bd}{\begin{displaymath}}
\newcommand {\ed}{\end{displaymath}}

\def \form#1 {eq. (\ref{#1}) }
\def \parziale#1#2  {{\partial {#1} \over \partial {#2}}}

\def\ec{e_{\mathrm{c}}}

\begin{document}

\title{On the origine of the Boson peak}

\author{Giorgio Parisi}

\address{Dipartimento di Fisica, Sezione INFN, SMC and INFM
 unit\`a di Roma 1, Universit\`a di Roma La Sapienza, Piazzale Aldo
 Moro 2, I-00185 Rome, Italy}

\begin{abstract}
We show that the phonon-saddle transition in the ensemble of generalized inherent structures (minima {\sl and} saddles)
happens at the same point as the
dynamical phase transition in glasses, that has been studied in the framework of the mode coupling approximation.  The
Boson peak observed in glasses at low temperature is a remanent of this transition.
\end{abstract}

\pacs{63.10.+a, 63.50+x, 64.70.Pf}

\maketitle

\section{Introduction}
{

The aim of this paper it show that the presence of a Boson peak is a necessity in the present days approach to glasses
(Angell 1995).  When we increase the temperature the inherent structures loose they stability: this happens at the
dynamical transition point, i.e. at the mode coupling transition $T_{c}$ (G\"oetze 1989).  In variance to what may
happens in other materials this lost of stability is due to the fact a finite part of the population of eigenvalues of
the harmonic spectrum migrates from the positive region to the negative region (Kurchan and Laloux 1996, Cavagna 2001,
Broderix \etal 2000, Angelani \etal).  In they journey these eigenvalues have to traverse the small $\omega$ region and,
when they do so, they produce the Boson peak.

The existence of the Boson peak it thus unavoidable and this explains its ubiquitous presence. However detailed computations
are needed to obtain its properties in quantitative way; in particular one would like to show that no anomalies in the
sound velocity are present at the frequency of the Boson peak. This paper presents some of the progresses that have 
recently done in this direction (Grigera \etal 2001, Grigera \etal 2002a, 2002b, 2002c)

In the next section of this paper, after the introduction, I will describe the general theoretic framework that stays
behind these computations.  In the next section I will introduce generalized inherent structures (minima and saddles) and 
I will 
discuss their properties.  In the fourth section I will compute in a simplified model the spectrum of the oscillations
around the generalised inherent structures and I will show how they are related to the Boson peak.  In the next
section I will show how more precise computations can be done in a more realistic models using the theory of Euclidean
random matrices.  In the last section I will present some conclusions. Finally in the appendix I will review some of 
the theoretical reasons that justify the relevance of the generalized inherent structures.

\section{The general framework}

It is usually believed that in the real world fragile glasses have only one thermodynamic transition 
with divergent viscosity (at temperature $T_{K}$).  This transition cannot be observed 
directly because the time required for thermalization is too long.  This transition is 
believed to be related to Kauzmann entropy crisis and it should happens just at the point 
where the configurational entropy becomes zero.  The viscosity is supposed to diverge as
$\exp(A/(T-T_{K}))$ and the specific heat should be discontinuous.

However in the idealized world of mean field theories, (Kirkpatrick \etal 1989), both in the framework of the mode
coupling theory (G\"oetze 1989) or of the equivalent replica approach (M\'ezard \etal 1987, Parisi 1992), if activated
processes are strictly forbidden (Franz and Parisi 1997), there is a second purely dynamics transition $T_{c}$ at higher
temperatures (for a recent review see Cugliandolo 2002, Parisi 2002).  Here the viscosity is divergent as a power law of
$T-T_{c}$.  This idealization is not so bad in the real world: activated processes are strongly depressed and the
viscosity may increase of many orders of magnitudo (e.g. 6) before reaching the region where activated processes become
dominant.

As it happens in many cases, slow relaxation is related to the existence of zero energy 
modes and this statement is true also in the mode coupling theory. This statement can be 
easily verified in spin models where the mode coupling theory is exact and simple 
computations are possible.

Summarizing, this qualitative description can be easily verified in models where the mean field 
approximation is exact.  In glass forming liquid, the picture is essentially sound 
(provided that we correct it by considering the existence of phonons). 

The aim of this note is to show that, if the previous picture holds, there is a Boson peak at low temperatures.  The
Boson peak is a bump in the density of vibrational states (divided by the Debye density of states that is proportional
to $\omega^{2}$) at low temperature in the low $\omega$ region (Benassi \etal 1996, Masciovecchio \etal, Sette \etal
1998, Engberg \etal 1999, Fioretto \etal 1999, H\'edoux \etal 2001). One of the remarkable and puzzling features of the
Boson peak (that is explained by present approach) is that the sound velocity is linear in the region of the Boson peak,
so that these low energy excitations do not appear at low momenta.

The Boson peak is present in many materials: an example of the experimental data in silica is given  in fig. \ref{BP}.

 \begin{figure} \begin{center}    
      \includegraphics[width=0.5\textwidth]{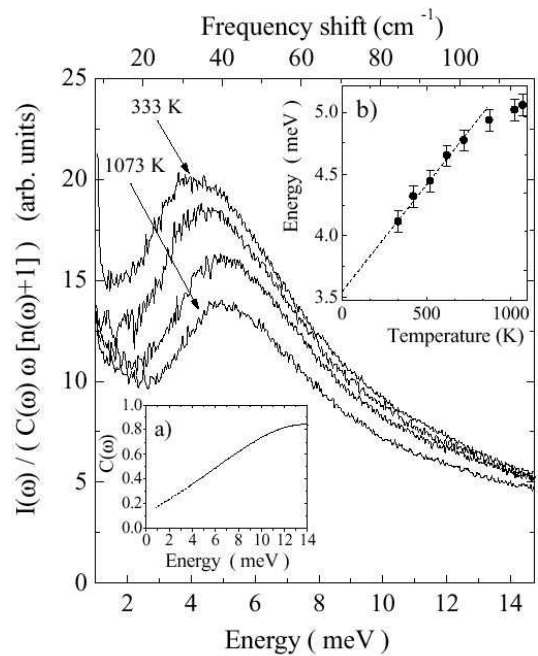}
      \end{center} \caption{ \label{BP}
 Examples of RS spectra taken in silica (Masciovecchio \etal). The  data correspond to the
reduced Raman intensities divided by the $C(\omega )$ 
 and shown in inset a). In inset b), the energy position of the maximum intensity at
each temperature ($\bullet $) is reported together with its linear fit in
the low temperature region.  
}\end{figure}

\section{Generalized inherent structures}

An inherent structure (IS) is a minimum of the Hamiltonian of the system that is near to a equilibrium configuration. 
(Stillinger 1995, Kob et al 2000, Debenedetti and Stillinger 2001).
We can associate to an equilibrium configuration an inherent structure as the nearest minimum.  In the same spirit  a
generalized inherent structure (GIS) is the nearest stationary point of the Hamiltonian (i.e. a point where the forces on 
all the particles are equal to zero (Cavagna 2001)).

It seems that with a very good approximation at temperatures lower than $T_{c}$ practically all GIS's are also IS's
(i.e. all stationary points of the Hamiltonian are also minima of the Hamiltonian) so that the two definitions
practically coincides in this region.  Only for $T>T_{c}$ starting from an equilibrium configuration the associated GIS
is not an IS and it has an higher energy (Broderix \etal 2000, Ruocco \etal 2000).

A relevant property of a GIS is its vibrational spectrum (and the associated  density of state);, if 
 $\lambda$ is an eigenvalue of the Hessian of the Hamiltonian, the frequency $\omega$ is given by
 \be
\omega\propto\sqrt{\lambda}\ . 
\ee
A crucial quantity is the fraction of negative eigenvalues (i.e. imaginary frequency) that will be denoted by $K$ .
If 
$K=0$ the GIS is an IS.

Generalized inherent structures are a powerful theoretical tool for many reasons.
\begin{itemize}
    \item At high temperature (i.e. at $T>T_{c}$) inherent structures are not relevant: they are quite far away from the
    equilibrium configurations.  On the contrary, also when you increase the temperature, you can find saddles that are not
    too different from the equilibrium configurations (Cavagna \etal 2002).
    \item Saddles are the natural continuation at higher temperature of the inherent structures 
    at low temperature and we may expect the the properties of the two different ensembles join smoothly at $T_{c}$.
     \item The most spectacular result is that the fraction of negative eigenvalues $K$ vanishes for the saddles when we
    approach $T_{c}$.  This property can be analytically proved in the framework of the mean field approach (Cavagna
    \etal 2002) and it is an empirical fact that it is satisfied with reasonable good approximation in all the models
    where explicit computations have been done (Broderix \etal 2000, Ruocco \etal 2002 , Grigera \etal  2002a).

 \end{itemize}
 
  \begin{figure} \begin{center}    
     \includegraphics[width=0.55\textwidth]{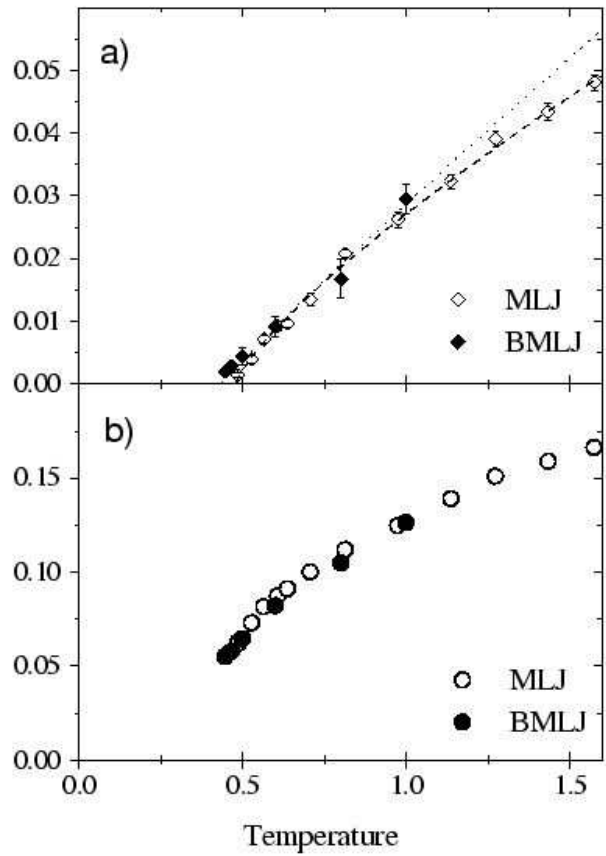}
       \end{center} \caption{Temperature dependence of the fraction of the negative eigenvalues of
the Hessian calculated at the inherent saddle configurations $n_s/3N$ (a),
and at the instantaneous configurations, $n_i/3N$ (b).
Open and closed symbols refer to two different choices of Lennard-Jones potentials (Angelani et al 2000).
\label{P2}}\end{figure}

The comparison of the spectrum of the instantaneous normal modes with that of the saddles is spectacular and is shown
in fig.  \ref{P2}.  The fraction of negative eigenvalues ($K$) of the saddles vanishes nearly linearly at $T_{c}$ while the
instantaneous normal modes do not display any interesting behaviour at this point. This behaviour of the fraction of 
negative eigenvalues of GIS's can be be used in many different ways.
\begin{itemize}
    \item The numerical computation of $K(T)$ can be used as a diagnostic tool to compute the value of $T_{c}$.
    \item
    The dynamical correlation time of the system (neglecting hopping) is divergent just at the point where $K(T)=0$.  It
    makes sense to try to relate in a quantitative way the properties of the spectrum around the saddles and the
    dynamical quantities (a similar effort would be hopeless for the instantaneous normal modes).  
    \item We can use the
    fact that the properties of the inherent structures at $T<T_{c}$ smoothly join with those of the saddles at 
    $T>T_{c}$ to predict the behaviour of the inherent structures at low temperature.
\end{itemize}

Here I will explore this third feature and I will show how one can derive in this framework the existence of a Boson 
peak.

\section{Exact non realistic computations of the spectrum}

It is well known that soluble models are not realistic and realistic models are not soluble. However the study of soluble 
models can give us some enlightens on the behaviour in a realistic model. This is particularly true in this case: 
many properties of the GIS's are very similar in realistic models and in the soluble cases.

The simplest model where we can investigate the properties of the GIS is the $p$-spin spherical model (Kirkpatrick \etal
1989).  This model is the most unrealistic one.  It has the advantage that, in spite of the fact that nearly everything
can be computed analytically, it has a quite rich behaviour (both $T_{c}$ and $T_{K}$ are well defined) and the mode
coupling equations for the dynamics are exact.

\begin{figure} \begin{center}    
\includegraphics[width=0.5\textwidth]{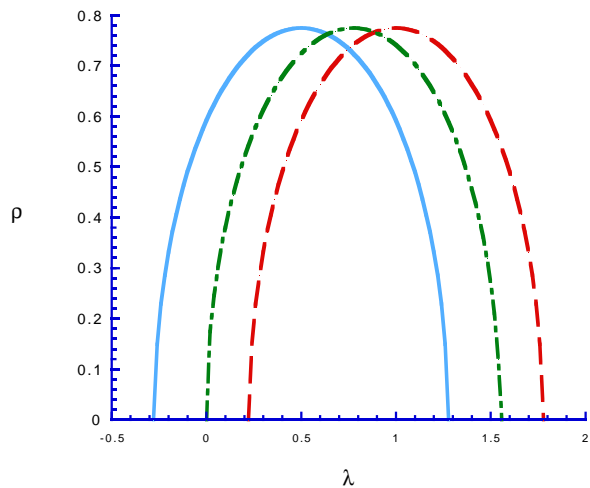}
       \end{center} \caption{
       The qualitative behaviour of the spectral density for the generalized inherent structures 
       in mean field approximation above 
    $T_{c}$, (full line), at $T_{c}$ (dot-dashed line) and below $T_{c}$ (dashed line) as 
    function of the eigenvalue $\lambda$).
\label{X}}\end{figure}

The spectrum of the harmonic oscillations around the inherent structures, the generalized inherent structures and the
instantaneous normal modes can be computed exactly (Biroli 1999, Cavagna \etal 2002): they have a semicircular shape
(see fig.  \ref{X}) whose edges are a function of the temperature.  The value of the lowest eigenvalue is particular
relevant for physics and it is shown in fig.  \ref{mu}.

If we look to spectrum of the oscillations around the inherent structures, we find that it has a gap at $T<T_{c}$ and 
this gap vanishes at $T_{c}$. Therefore at the dynamical transition there is an excess of low frequency modes
with respect to what happens at lower temperatures. This result maybe unexpected; however it is a necessary consequence of 
the fact that at $T_{c}$  the inherent structures merge with the saddles  and the saddles must have negative eigenvalues.

Skipping all the details we conclude that there is a population of modes whose eigenvalues decrease when we
approach $T_{c}$ and these eigenvalues change sign when we cross $T_{c}$.  This is a quite general phenomenon and 
it survives also in more realistic cases. It is clear that the modes that migrate at low temperature must  produce  
an increase in the density of states at low energy with respect to what happens in other models where these modes are 
non present (e.g. a crystal).

It is quite natural  to suppose that this excess of extra modes  is the origine of the Boson peak. In order to prove the 
correctness of this intuition there are three crucial points that must be discussed:
\begin{itemize}
    \item This effect  produces only an excess of modes in the low momenta region or do we find a peak, after dividing 
    by the Debye density?
    \item Do these modes produce an anomaly in the sound velocity, and if not, why?
    \item
    The Boson peak is something that appears at low temperatures, the effect I am speaking about happens near $T_{c}$.
    How can one use a property of the system near $T_{c}$ to deduce the behaviour in the low temperature region?
\end{itemize}

It order to be able to answer to the previous questions it is necessary to do some explicit computation in more realistic 
3-dimensional models and this will be done in the next section.

 \begin{figure} \begin{center}    
     \includegraphics[width=0.5\textwidth]{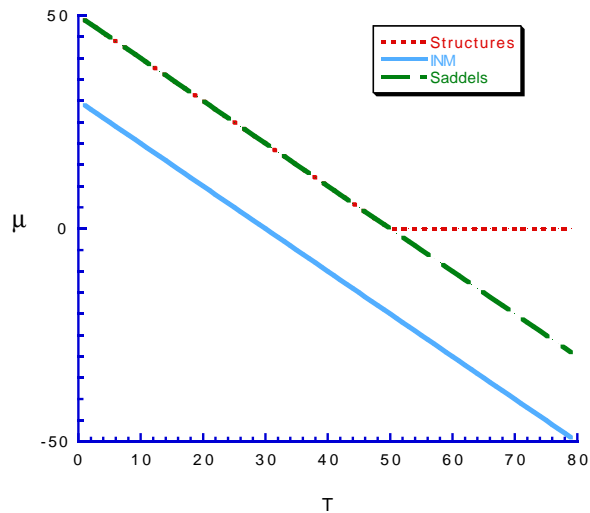}
      \end{center} \caption{
      The qualitative behaviour of the threshold as function of the temperature
    in mean field approximation for the generalized inherent structures  normal modes (dotted line), for the 
    instantaneous normal modes (full line) and for the inherent structures normal modes (dashed line).
\label{mu}}\end{figure}
   
\section{Euclidean Random Matrices and the phonon-saddle transition}

Generally speaking our task is to compute the spectrum of the second derivative of the Hamiltonian, i.e. the spectrum 
of the Hessian the matrix $M$ defined as
\be
M_{i,k}=
\delta_{i,k} \sum_{j}V''(x_{i}-x_{j}) -V''(x_{i}-x_{k}) \ . 
\ee
The points $x$ are chosen random with a given probability distribution.
\begin{itemize}
\item If the points $x$ are extracted with the equilibrium probability distribution  at a temperature $\beta$, we get the 
instantaneous normal models.
\item If the  points $x$  belong to a random GIS, we get the spectrum of the GIS.
\item If the points  $x$  belong to a random IS, we get the spectrum of the IS.
\end{itemize}

We can  generalize the problem by studying the behaviour of the spectrum of $M$ when we change the probability 
distribution of the points $x$'s.  For general random distribution we have the so called problem of Euclidean random 
matrices (M\'ezard \etal 1999, 2001).

In general we can stay in one of the two phases: all the eigenvalues of $M$ are positive (phonon phase), a fraction of
the eigenvalues of $M$ are negative (saddle phase).  By changing the parameters of the distribution of $x$ we should go
from one phase to an other one \footnote{A technical remark: in many ensembles the matrix $M$ has a tails of localized
eigenvalues that may extend up to infinity or very far away.  In this situation the phonon phase is impossible.  In the
following we are going to neglect the existence of these localized modes whose fraction is often very small.  If we take care
of the existence of localized modes the phonon-saddles transition is no more exactly sharp.  This is not a surprise
because also the dynamical transition $T_{c}$ is not defined with infinite precision.  The dynamical transition becomes
sharp if activated processes are neglected and the phonon-saddle transition become sharp if localized modes (in the low
part of the spectrum) are neglected.  These two approximations are related but I cannot discuss more this point for
reasons of space.}.

This approach can be successful only if we are able to put under control the general problem of Euclidean random models.
The strategy for the study of the spectrum of these random operators consists in extending to  topologically disordered
systems the approach used in disordered lattice problems. 

The first step consist in extending the usual CPA approximation of lattice systems to the present case. This can be 
done in a few steps (Grigera \etal 2001, 2002a, 2002b, 2003):

\begin{enumerate}  \item One firstly identifies a soluble limit (jellium) where the density of the particles is very high 
or equivalently the range of the forces goes to infinite.

\item The corrections to the jellium limit can be computed in a systematic way  as expansion in powers of the inverse 
of the density. This expansion can be expressed in a diagrammatical way.

\item It is possible to resum a given class of diagrams (very similar to those of the CPA) and to arrive to some
kind of integral equation of the form
\be G(p,\omega) ^{-1}=G_{0}(p,\omega)+ \int dk \, W(p,k)^{2} G(p,\omega)
\ee
where $G$  is the Green function \footnote{The Green function is related to the usual structure function by the
relation: $ S(p,\omega)\propto p^{2} \omega^{-1} \mbox{Im} \ G(p^{2},\omega+i 0^{+}) $.}  and it is equal to 
the average of the resolvent, i.e.
\be
G(p,\omega)=\overline{\sum_{j,k} R(j,k|\omega^{2}) \exp(i p(x_{j}-x_{k})} \ ,
\ee
and the resolvent is given by 
\be
R(j,k|\lambda)\equiv \left({1 \over \lambda -M }\right)_{j,k} \ .
\ee

\item These integral equations can be numerically solved.  For a suitable choice of the parameters one
find a phonon saddle transition \footnote{The leaned reader who is concerned by the fate of localized states, should 
notice that localized states are not be present in this CPA approximation.}
Fortunately is also  possible to derive  some analytic results near transition and to obtain the value of critical exponents..
\end{enumerate}

Numerical and analytic studies show that there is an anomaly in the spectrum that has a shape very similar to the 
Boson peak. This has been done for unrealistic potential and the computation for realistic potential is going on and 
should be ready soon (Grigera \etal 2002). 

The main result (that was  unexpected, at least by me) is that near the transition nothing happens in the small
momentum region.  The sound velocity is regular in the region of frequency  of the Boson peak. The  states relevant for 
the transition have simultaneously 
high momentum and  low energy region; there is a only marginal hybridization of these states 
with the acoustic branch (however this hybridization influences the value of the critical exponents).

It is quite possible  that high order correction do not affect the  critical exponents; however this system behaves in a 
very different way from the other phase transitions that I have studied in the last thirty years and I can hardly make 
a reliable guess before further investigations.

It is reasonable to suppose that these properties are universal and they are also present in the phonon-saddle transition
for the generalized inherent structures as function of the temperature.  However we must remark that the relevant
parameter on which we expect a smooth behaviour is not the initial temperature, but the energy of the final generalized
inherent structure: when the temperature change from $0$ to $T_{K}$ the properties of the generalized inherent
structures (energy and spectrum) should hardly change, they should change by a little amount when we go from $T_{K}$ to
$T_{c}$ and they should strongly change when we go from $T_{c}$ to a very large temperature.  In other words, due to the
fact that the ration $T_{c}/T_{K}$ is not very far from 1, the inherent structures should change of a little amount
when we go from $T_{c}$ to $0$.  In other word the energy of the IS structure at zero temperature is not very different
from the energy of the IS at the critical temperature.  In this sense 0 temperature is near to $T_c$ and therefore the
behaviour of the system  in the critical region is relevant also at low temperature.

More detailed predictions can be done, but I cannot discuss them for reasons of space.

\section{Conclusions}

Let me try to summarize in a very crude way the  scenario that is emerging.

In pure mean field models the excitation around a generalized saddle point form a band of delocalized states (let us 
call then glassons in absence of  a better name). 
When we cross the dynamical transition the lower edge of the glassons goes from negative eigenvalues (imaginary frequency)
to positive eigenvalues (real frequency).

In realistic models we have both the glasson band and the usual phonon band. The low momenta structure function is 
dominated by the phonons and glassons decouple in this region. As far as the two bands superimpose it is not possible 
to separate them in a sharp way because there is always an hybridization among them, however it is convenient to think of 
them as two separate entities.  At the dynamical point the lower edge of the glasson band becomes zero.  At the
low temperatures the glasson band develops a gap and when we look to the total density of states divided by $\omega^{2}$,
the opening  of the glasson band shows up as the Boson peak. The {\sl small} hybridization among phonons and 
glassons is crucial to determine the detailed behaviour of the Boson peak near the dynamical transition.

The main conclusion of this analysis are the following:
\begin{itemize}
\item The Boson peak is a remanent of the softening of the free energy landscape at the dynamical temperature
and it composed  by modes that migrate at imaginary $\omega$ at $T>T_{c}$.

\item If activated processes were suppressed in the dynamics, the Boson peak would be infinite.  In the same vein
ultrafast quenching of the sample should give a very strong dependance of the Boson peak on the temperature (see Angell's
contribution to this conference).

\item In general a Boson peak is present in any system  of matrices near a phonon-saddle transition. In the case of the 
generalized inherent structure the point where this phenomenon happens coincide with the dynamical transition.

\item Quantitative analytic direct computations of the various properties of the Boson peak are feasible and they will
be done a next future.

\item A comparison of this fully microscopic approach with the results of the mode-coupling theory should
be possible (G\"oetze and Mayr 2000): work in this direction is in progress.

\end{itemize}

There still are many unclear points, conjectures that must be verified, connections with other theoretical results that must 
be established, but I believe that the basic scenario has been drawn and that it is essentially correct. 

\section*{Appendix}
Here I would like to describes some of the reasons  for which generalized inherent structures are important.

The general idea is quite simple.  Let us consider the free energy as functional of the density $\rho(x)$ (i.e.
$F[\rho]$).  We expect that at $T>T_{c}$ the trivial solution $\rho(x)=const$ is the only solution of the stationary
equations the free energy that is relevant for the thermodynamics.  At low temperatures there are an exponential large number
of non-equivalent solutions where the the density has a non trivial dependence on $x$.  Skipping many details the
situation should be the following:

\begin{itemize}
    \item A temperature $T>T_{c}$ there are no non-trivial relevant solutions of the equation 
    \be
    {\delta F \over \delta \rho(x)}=0, \label {D1F}
    \ee
    however the dynamics is dominated by quasi solutions of the previous equations, i. e. 
    by densities $\rho(x)$ such that the left hand side of the previous equation is not 
    zero, but vanishes when $T$ approach $T_{c}$. These quasi solutions are relevant for 
    the dynamics (Franz and  Virasoro 2001). One can compute the spectrum of the Hessian
    \be
    M(x,y)={\delta^{2} F \over \delta \rho(x) \delta \rho(y)}\label {D2F} \ .
    \ee
    One finds that  $M$ has negative eigenvalues and its spectrum extends to
    the negative eigenvalue region and has qualitatively the shape shown in fig. \ref{X}.
    These quasi-stationary points of $F$ look like saddles.
    \item
    At the transition point $T=T_{c}$ the quasi stationary points becomes real solutions 
    of the equations (\ref{D1F}). They are essentially minima: the spectrum of the Hessian 
    is non-negative and it arrives up to zero. As it can be 
    checked directly, the existence of these nearly zero energy modes is responsible of 
    the slowing down of the dynamics. The different minima are connected by flat regions 
    so that the system may travel from one minimum to an other (Kurchan and Laloux 1999).
    \item
    At low temperature the mimima become more deep, the spectrum develops a gap as shown 
    in fig. \ref{X} and the minima are no more connected by flat regions. If activated 
    processes were suppressed, the system would remains forever in one of these minima. In 
    the real world the system may jump  (by decreasing his energy) until it reaches the 
    region where the minima are so deep that the energy barriers among them diverges.
\end{itemize}

This picture is not so intuitive because it involves the presence of saddles with many 
directions in which the curvature is negative, and it is practically impossible to 
visualize it by making a drawing in a two or a three dimensional space.

This qualitative description can be easily verified in models where the mean field 
approximation is exact.  In glass forming liquid, the picture is essentially sound 
(provided that we correct it by considering the existence of phonons).  However, if we try 
to test it a more precise way, we face the difficulty that the free energy functional 
$F[\rho]$ is a mythological object whose exact form is not exactly known and consequently the 
eigenvalues of its Hessian cannot be computed.  

The generalized inherent structures are the ``poor man'' substitute of the solutions (or quasi solutions) of the 
stationary equations of the free-energy. It can  be checked in mean field theory (Cavagna et al 2002) that the two 
constructions are quite similar and it is therefore reasonable that this similarity remains true also in more realistic 
models. It remains however surprising how the whole picture can be transferred from mean models to realistic models 
without too much to change.

\section*{Acknowledgment}I am  happy to thank  A. Cavagna, I. Giardina, T. Grigera, V. Mart\'\i n-Mayor and 
P. Verrocchio who have worked with me on these matters in the last two years
\section*{References}
\begin{harvard}
    
\item Angelani, L., Di Leonardo, R., Ruocco, G., Scala, A., and Sciortino, F., 2000 {\em Phys. Rev. Lett.} {\bf 85,} 
5356--5359. 
\item Angell, C. A., 1995 {\em Science} {\bf 267,} 1924--1935 
\item Benassi, P. \etal 1996 {\em Phys.  Rev.  Lett.} {\bf 77,} 3835--3838. 
\item Biroli G. 1999, J. Phys. Math. Gen.  {\bf 32} 8365. 
\item Broderix, K., Bhattacharya, K. K., Cavagna, A., Zippelius, A., and Giardina, I., (2000) {\em Phys.  Rev.  Lett.} 
5360.
\item Cavagna, A (2001), Europhys. Lett. {\bf 53}, 490. 
\item Cavagna A. , Giardina I, and Parisi G. (2002), Phys, Rev. Lett.
\item Cugliandolo L.F., (2002), Lectures given to the Les Houches summer School 2002, (in press). 
\item DeBenedetti, P. G., and Stillinger, F. H., (2001).  {\em Nature} {\bf 410,} 259--267. 
\item Engberg, D., \etal (1999) {\em Phys.  Rev.  B} {\bf 59,} 4053--4057. 
\item Fioretto, D., \etal 1999 {\em Phys.  Rev.  E} {\bf 59,} 
4470--4475 . 
\item Franz S. and Parisi G. , 1997, Phys.  Rev.  Letters {\bf 79}, 2486. 
\item Franz  S. and Virasoro M.A. 2000, J. Phys.  A: Math.  and Gen.  {\bf 33}, 891.
\item G\"oetze W. 1989 {\em Liquid, freezing and the Glass transition}, Les Houches, J. P. Hansen, D. Levesque, J.
Zinn-Justin editors, North Holland. 
\item G\"oetze W., and Mayr M., (2000) {\em Phys.  Rev.  E} {\bf 61,} 587--606
\item Grigera, T. S., Mart\'\i{}n-Mayor, V., Parisi, G., and Verrocchio, P. (2001).  {\em Phys.  Rev.  Lett.} {\bf 87,} 
085502-1--085502-4 
\item Grigera, T. S., Cavagna, A., Giardina, I., and Parisi, G., 2002a {\em Phys.  Rev.  Lett.} {\bf 88,} 
055502-1--055502-4 .55 
\item Grigera, T. S., Mart\'\i{}n-Mayor, V., Parisi, G., and Verrocchio, P. (2002b)., {\em J. Phys.: Condens. Matter} {\bf 14,} 
2167--2179. 
\item Grigera, T. S., Mart\'\i{}n-Mayor, V., Parisi, G., and Verrocchio, P. (2002c),  in preparation.
\item Grigera, T. S., Mart\'\i{}n-Mayor, V., Parisi, G., and Verrocchio, P. (2003), cond-mat/0301103,  
to be published on Nature.
\item H\'edoux, A., Derollez, P., Guinet, Y., Dianoux, A. J., and Descamps, M., (2001).  {\em Phys.  Rev.  B} {\bf 63,} 
144202-1--144202-8 . 
\item Kirkpatrick T. R. , Thirumalai D.  and Wolynes P.G. 1989,  Phys. Rev. {\bf
A40}, 1045. 
\item Kob, W., Sciortino, F., and Tartaglia, P., 2000, {\em Europhys. Lett.} {\bf 49,}590--596.
\item Kurchan J. and Laloux L. (1996), J. Phys. A {\bf 29}, 1929. 
\item M\'ezard, M. Parisi G. and Virasoro M.A. 1987 {\sl Spin
glass theory and beyond}, World Scientific (Singapore)  
\item M\`ezard, M., Parisi, G., and Zee A., (1999). {\em Nucl. Phys. B} {\bf 559,} 689--701.
\item M\`ezard, M., Parisi, G., and Verrocchio P. 2001, J.\ Chem.\ Phys.\ {\bf 114}, 8068.
\item Mart\'\i n-Mayor, V., M\`ezard, M., Parisi, G., and Verrocchio, P., (2001).  {\em J. Chem.  Phys.} {\bf 114,} 8068--8081
\item Masciovecchio, C. \etal, 1996.{\em Phys.  Rev.  Lett.} {\bf 76,} 3356--3359 (1996). 
\item Parisi G. (1992), {\sl Field Theory, Disorder and Simulations}, World Scientific, (Singapore ). 
\item Parisi G, (2002), Lectures given to the Les Houches summer School 2002, (in press). 
\item Ruocco, G.  (2000) \etal  {\em Phys. Rev. Lett.} {\bf 84,} 5788--5791. 
\item Sette F., Krisch M. H., Masciovecchio C., Ruocco G., and Monaco G., (1998){\em Science } {\bf 280,} 1550--1555. 
\item Stillinger, F. H. 1995, {\em Science} {\bf 267,} 1935--1939,
{\bf 85,} 5360--5363 .

\end{harvard}
\end{document}